\documentclass[12pt,letterpaper]{article}
\pdfoutput=1
\usepackage{jheppub}
\usepackage{amsmath,amssymb,scalefnt}
\usepackage{graphicx}
\usepackage{caption}
\usepackage{subcaption}
\newcommand{\beq}{\begin{equation}}
\newcommand{\eeq}{\end{equation}}
\newcommand{\bea}{\begin{eqnarray}}
\newcommand{\eea}{\end{eqnarray}}
\newcommand{\nn}{\nonumber\\}

\def\pa{\partial}
\newcommand{\vev}[1]{\left\langle#1\right\rangle}
\newcommand{\eqn}[1]{Eq.~(\ref{#1})}
\newcommand{\eqns}[2]{Eqs.~(\ref{#1}),(\ref{#2})}

\newcommand{\reference}[1]{Ref.~\cite{#1}}

\newcommand{\half}{\frac{1}{2}}

\author[a,1]{Martin B Einhorn}\note{Also, \it{Michigan Center for Theoretical Physics, Ann Arbor, MI 48109.}}
\author[a,b]{and D R Timothy Jones}
\affiliation[a]{Kavli Institute for Theoretical Physics,\\ University of California, 
Santa Barbara, CA 93106-4030, USA}
\affiliation[b]{Dept. of Mathematical Sciences,\\ University of Liverpool, Liverpool L69 3BX, UK}

\emailAdd{meinhorn@umich.edu}
\emailAdd{drtj@liv.ac.uk}

\title {Induced Gravity I: Real Scalar Field}  

\abstract{We show that classically scale invariant gravity coupled to a
single scalar field  can undergo dimensional transmutation and generate
an effective Einstein-Hilbert action for  gravity, coupled to a massive
dilaton. The same theory has an ultraviolet fixed point  for coupling
constant ratios such that all couplings are asymptotically free. However
the catchment basin of  this fixed point does not include regions of
coupling constant parameter space compatible with locally stable
dimensional transmutation.  In a companion paper, we will explore whether
this more desirable outcome does obtain in more complicated theories
with non-Abelian gauge interactions.}


\begin{document}
\maketitle

\section{Introduction}\label{sec:intro}

The framework for this paper is classically scale invariant
quantum gravity,  defined by the Lagrangian   
\beq\label{eq:hoaction} 
S_{ho}=\int d^4x\sqrt{g}\left[ \frac{C^2}{2a}+\frac{R^2}{3b}+cG \right],
 \eeq 
where $R$ is the Ricci scalar, $C$ is the Weyl tensor and $G$ is the 
Gauss-Bonnet (GB) term. There are three 
dimensionless coupling constants, $(a,b,c)$.   
Just about the
simplest imaginable scale invariant theory involving gravity and matter
fields   consists of  the above, coupled to a single scalar field with a
$\lambda\phi^4$ interaction and  non-minimal gravitational coupling $\xi
R \phi^2$.
In a recent paper~\cite{Einhorn:2014gfa}, we  argued that even this
basic theory can undergo dimensional transmutation (DT)  \` a la  
Coleman-Weinberg (CW)~\cite{Coleman:1973jx}, leading to effective action
extrema with  nonzero values of the curvature 
and of the scalar field.\footnote{Some early work in the same spirit, 
but in the special case of a {\it conformal\/} theory, can be found in 
\reference{Buchbinder:1986wk}; see also \reference{Cognola:1998ve} and references therein.  Whether the conformal version of this theory is renormalizable remains controversial.  See, e.g., \reference{Shapiro:1994ww}.}
It is important to emphasise that, as in the original CW treatment  of
massless scalar electrodynamics,  we restrict ourselves to DT that can
be demonstrated perturbatively,  in other words, for values of the
relevant dimensionless couplings such that higher-order quantum
corrections are small.

In this paper, we revisit the results
of~\reference{Einhorn:2014gfa},  while in a companion paper~\cite{EJ}\
we extend our approach
to the case  when the matter sector includes gauge
interactions  and matter fields with a more complicated scalar sector.  
Our goal in this will be  to demonstrate that the same DT process can be
responsible for generating  both the Planck mass (with the associated
gravitational interactions)  and  the breaking of a Grand Unified gauge
symmetry. In addition we seek a theory such that all dimensional
couplings are  asymptotically free (AF), with the region of DT within
the basin of attraction  of an ultra-violet stable fixed point
(UVFP) for ratios of couplings.  In the case of the minimal model
treated here, this is {\it not\/} the case;  although a UVFP {\it does\/} 
exist, with all the dimensionless couplings AF, the DT region is 
not within its catchment basin.

Before we proceed to gauge theories, however, 
we have to reassess  our previous calculations, for the
following reason. Critical to the demonstration of DT  in these theories
are the results for the one-loop beta-functions,  including those of
the gravitational self interaction couplings, as well  as the
contributions of these couplings to the  beta-functions for the
matter interactions.  These were calculated some  time
ago~\cite{Fradkin:1981hx, Fradkin:1981iu, 
Avramidi:1985ki, Avramidi:2000pia, Buchbinder:1992rb}
and were
summarized in \reference{Buchbinder:1992rb} (BOS) for a range of
theories. Calculations of this type were revisited recently by Salvio
and Strumia~\cite{Salvio:2014soa} (hereafter, SS),  with results differing
significantly from the earlier ones for the beta-functions  for the 
interactions involving matter fields.\footnote{There is no change to the 
gravitational coupling beta-functions (see \reference{Avramidi:1985ki}).} 
We shall see, however, that using the correct 
beta-functions does not alter the essential conclusions of 
\reference{Einhorn:2014gfa}.

While we endorse the SS form of the beta-functions in general, we differ from 
them in one respect that impacts the DT calculation. They 
rewrite $C^2$ as follows 
\beq
C^2 = G + 2W
\eeq
where $W = R_{\mu\nu}^2 -\frac{1}{3}R^2$, so that \eqn{eq:hoaction}\ becomes 
\bea\label{eq:SShoaction} 
S_{ho}&=&\phantom{-}\int
d^4x\sqrt{g}\left[ \frac{1}{a}{W}+\frac{R^2}{3b}+(c+\frac{1}{2a})G \right]\nn
&=&-\int
d^4x \sqrt{g}\left[ \frac{1}{f_2^2}(\frac{1}{3}R^2-R_{\mu\nu}^2)
+\frac{R^2}{6f_0^2}
-(c+\frac{1}{2a})G \right],\eea
where $a=f_2^2$ and $b = -2f_0^2$, {\it and then ignore the $G$ term  
throughout}, on the grounds that  it can be expressed locally as a total derivative. The problem 
with this strategy, and one specifically relevant to the DT paradigm, 
is that the theory without the $G$ term is not 
multiplicatively renormalisable~\cite{Fradkin:1981iu}. In curved space but with gravity 
not quantised, the beta-function associated with the renormalisation of $G$ 
is the Euler anomaly coefficient; for a recent discussion of its generalisation 
to the quantised gravity case considered here, see \reference{Einhorn:2014bka}.
The beta-function for the coefficient of the GB term enters the 
equation for DT, 
to be discussed in Section~\ref{sec:DT1} below.

We also differ from SS in that we conclude (as before~\cite{Einhorn:2014gfa}),
that we require both $a > 0 $ and $b > 0$,  whereas they claim that 
there is a tachyonic mode  if $b > 0$
(corresponding to $f_0^2$ negative in SS). We will discuss this issue further 
in section~\ref{sec:MM}.

\section{Fixed points and asymptotic freedom}
\label{sec:FP}
One attractive property of pure renormalizable gravity is that it is 
asymptotically free (AF)~\cite{Fradkin:1981hx, Fradkin:1981iu},
and this property  can be extended to include
a matter sector with an asymptotically free gauge theory or even a
non-gauge theory. 
This can be seen as follows.   At one-loop order, a gauge coupling
$g^2$ and the couplings $a$ and $c$ do not mix with the other couplings.
 In the general case, their beta-functions are 
(we suppress throughout a factor 
$(16\pi^2)^{-1}$ from all one-loop beta-functions):
\begin{subequations}
\label{eq:betagac}
\begin{align}
\beta_{g^2}&=-b_g\, (g^2)^2, &\qquad \beta_a&=-b_2\, a^2,&\qquad \beta_c&=-b_1,\\
\label{eq:bgb2b1}
b_g&=2(\frac{11}{3}C_G -\frac{2}{3}T_F -\frac{1}{6}T_S),  &\qquad 
b_2&= \frac{133}{10}+ N_a,  &\qquad 
b_1&=\frac{196}{45}+ N_c,
\end{align} 
\end{subequations}
where $N_a = \left[N_0+3N_F+12N_V\right]/60$ and 
$N_c =\left[N_0+\frac{11}{2}N_F+62N_V\right]/360$. 
Here $N_0$, $N_F$ and $N_V$ are the numbers of (real) scalar,
(two-component) fermion,  and (massless) vector fields respectively.
(Note that $N_F = 2N_{\half}$, the number of fermions  as defined
in~\reference{Einhorn:2014gfa} and earlier works.) $C_G$, $T_F$ and $T_S$
are the usual  quadratic Casimirs for the pure gauge theory and fermion
and scalar representations, with  the coefficients of $T_F$ and $T_S$ in 
\eqn{eq:bgb2b1}\ also reflecting our choices
of two component fermions and real scalars. 

It is worth noting at this point that whereas $g$ is AF for $b_g > 0$ 
whether it is positive or negative (the sign of the gauge coupling 
is not a physical observable), for $a$ to be AF we must 
have $a > 0$; $a < 0$ corresponds to an unphysical 
phase with a Landau pole in the UV. 
Similarly the coupling $c$ is asymptotically free for $c > 0,$ 
since $b_1 > 0$.

The evolution of $b$ is more complicated, because  $b$ mixes with the
couplings $a,\xi$; moreover,  $\beta_{\xi}$ depends  on the matter
self-couplings.  Therefore  the evolution of $b$   must be discussed
model-by-model. (Note, however,  that  all three purely gravitational
couplings  $(a,b,c)$ have  beta-functions independent of the gauge
couplings (if any) at one loop.) Clearly the possibility of completely
AF theories exists for  non-gauge theories and for  non-abelian gauge
theories,  but never for an abelian gauge coupling.   Thus, the models
of interest cannot have gauged U(1) factors, contrary to much of the
landscape of string theories.

In a certain sense, the evolution of the two couplings   $a$ and $g^2$
control the behavior of the other couplings in the theory. To see this,
it is useful to rescale the other couplings by one of these two and to
express their beta-functions in terms of these ratios.  In theories
without AF gauge couplings, one must choose $a,$ as we did in our
previous papers.   In gauge  models, it is more
convenient~\cite{Buchbinder:1992rb}  to rescale by $g^2$ instead, 
replacing the conventional running parameter $dt=d\ln\mu$ by $du=g^2(t)
dt.$  

\section{The Minimal Model}
\label{sec:MM}

The Minimal Model as described in \reference{Einhorn:2014gfa} consists 
of the action  
\beq S = S_{ho} + S_{\phi}, 
\eeq 
where
\beq\label{eq:jrealscalar} 
S_{\phi}=\int d^4x \sqrt{g}\left[ \half
(\nabla\phi)^2+\frac{\lambda}{4}\phi^4-\frac{\xi\phi^2}{2}R \right].
\eeq 
Our analysis proceeded in two stages; determination of the fixed point structure 
of the RG evolution, and demonstration of the existence of extrema 
determined by DT. 

\subsection{The Fixed Points}

The relevant beta-functions are $\beta_{a,c}$ from 
\eqn{eq:betagac}, 
and $\beta_{b, \lambda,\xi}$ given by 
\beq
\label{eq:betab}
\beta_b\equiv -a^2 b_3(x,\xi),\quad  
b_3(x,\xi)\equiv \left[\frac{10}{3}-5\,x+ \left(\frac{5}{12} 
+\frac{(6\xi+1)^2}{24}\right)x^2\right],\\
\eeq
where $x \equiv b/a,$ 
\beq\label{eq:betalambda}
\beta_{\lambda} =18\lambda^2 +\half \xi^2(5a^2 +\frac{1}{4}(6\xi+1)^2b^2)
+\lambda(5a-\frac{1}{2}(1+6\xi)^2 b),
\eeq
and 
\beq\label{eq:betaxi}
\beta_{\xi}=(6\xi+1)\lambda +\frac{1}{3}\xi\left(\frac{10a^2}{b}
-(9\xi^2+\frac{15}{2}\xi+1)b\right).
\eeq
As described in the introduction, the results for $\beta_{\lambda,\xi}$ above 
correspond to those 
of SS and differ significantly from those employed by us in 
\reference{Einhorn:2014gfa}\footnote{In comparing with SS, 
one must bear in mind that they 
use a {\it complex} scalar singlet.}, based on the beta-functions in the earlier 
literature~\cite{Buchbinder:1992rb}.
For example, although there is a $b\xi^3$ term in \eqn{eq:betaxi}, 
there is no $a\xi^3$ term;  and it is easy to show by an expansion 
of the metric about flat space and by consideration of 
the respective contributions of the $a$ and $b$ terms to the graviton propagator
that that no such term can arise. In a similar way it can be shown that there can be  no 
$\lambda a\xi^2$ term in $\beta_{\lambda}$. However, such terms appear in the 
expressions corresponding to \eqns{eq:betalambda}{eq:betaxi}\ in
BOS, which is one reason we believe them to be incorrect.

To determine the asymptotic behavior of the couplings 
for the above system of beta-functions, we make a 
couple of redefinitions. 
We introduce $x = b/a$, (more convenient than $w=a/b$ employed in 
\reference{Einhorn:2014gfa}, it turns out) and $y=\lambda/a$, and a 
running parameter $u$ such that $du = a(t)dt$.

We then obtain the reduced set of beta-functions:
\begin{subequations}
\label{eq:betabarmat}
\begin{align}
\frac{dx}{du}\equiv\overline{\beta}_x
&=-\frac{10}{3}\left[
1-\frac{1099}{200}x+\frac{1}{8}x^2
+\frac{1}{80}\left(1+6\xi\right)^2 x^2 \right]; 
\label{eq:betabarmatA}\\
\frac{d\xi}{du}\equiv\overline{\beta}_{\xi}&=\left(6\xi 
+ 1\right)y +\frac{\xi}{6}\left(\frac{20}{x}-x(6\xi+1)(3\xi+2)\right);
\label{eq:betabarmatB}\\
\label{eq:betabarmatC}
\frac{dy}{du}\equiv\overline{\beta}_y&=18y^2 
+ y\left(\frac{1099}{60}-\half x (1+6\xi)^2\right)
+
\frac{\xi^2}{8}(20+(6\xi+1)^2x^2).
\end{align}
\end{subequations}
Now from \eqn{eq:betabarmatA}\ it is easy to show that 
FPs can only exist for $-4.23 \leq \xi \leq 3.89$, and that for 
values of $\xi$ in this range $\overline{\beta}_x = 0$ has 
two solutions for $x$, both with $x > 0$, 
with the smaller and larger values of $x$  being 
IR and UV attractive respectively. 

The fixed points of this system  of beta-functions  (and their
nature) are given in  Table~1. Remarkably, one of the fixed points with
$y=\xi=0$ is  UV stable (it is easy to see that a FP with $y=0$ must
have $\xi = 0$).  Since $a$ is AF,  this FP corresponds to AF for all
the couplings  $(a,b,c,\xi,\lambda)$. With regard to the IR stable FP, 
note that in approaching it from any starting values of the couplings, one would eventually 
lose perturbative believability since in the IR the coupling $``a"$ approaches a Landau 
pole in this limit. 

We will explore later the catchment basin of the UVFP; but note that
since $x>0$ at the FP,  it is clear from \eqn{eq:betabarmatA} that no
region of parameter space with $x < 0$ lies in this basin.
(Manifestly, for $x < 0$, $\overline{\beta}_x < 0$ as well, so $x\to-\infty$ starting at any value of $x<0$.) Thus, since we have already 
concluded that $a > 0$, we must have $b > 0$ as well at any scale 
from which the couplings can possibly 
approach the UVFP at higher energies.

\begin{table}[ht]
\begin{center}
\begin{tabular}{|c|c|c| c| c|  } \hline
& $ x $ & $ \xi $ & $ y $ & Nature\\ \hline
$\!{\bf 1.} $ & ${\bf 39.78082} $ & $ {\bf 0} $
& $ {\bf 0} $ & {\bf UV stable} \\ \hline
$2.\ $ & $ 0.18282 $ & $ 0 $ & $ 0 $ & IR stable\\ \hline
$3.\ $ & $ 0.18292 $ & $ 0.083150 $ & $ -1.005218$ & saddle point\\  \hline
$4.\ $ & $ 36.9666 $ & $ 0.058999$ & $ 0.787391 $
& saddle point \\ \hline
$5.\ $ & $ 43.7762 $ & $ -0.16404 $ & $ -1.01350 $ & saddle point \\ \hline
$6.\ $ & $ 43.7770 $ & $ -0.16551 $ & $ -0.0037756$ & saddle point \\  \hline
\end{tabular}
\caption{\label{FP}Fixed Points}
\label{table:FP}
\end{center}
\end{table}
\subsection{Dimensional Transmutation}\label{sec:DT1}

In  \reference{Einhorn:2014gfa} we discussed this theory in a totally 
symmetric gravitational background: 
\beq
R_{\mu\nu\lambda\rho} = \frac{R}{12}\left(g_{\mu\lambda}g_{\nu\rho} 
- g_{\mu\rho}g_{\nu\lambda}\right),
\eeq
when the classical action $S_{cl}$ can be written  
\beq 
\frac{S_{cl}}{V_4} =
\frac{1}{3b} + \frac{c}{6}  + \frac{1}{4}\left[ \lambda r^2-2\xi r
\right], 
\eeq 
where $V_4$ is a dimensionless volume element independent of $R$,
(we will rescale $S_{cl}$ to absorb $V_4$ henceforth) and  $r \equiv
\phi^2/R$.  The action has an extremum for $r= r_0 = \xi/\lambda,$ which
is a local minimum if $\lambda>0,$ where it takes the value
\beq\label{eq:sonshell}
S_{os}=\frac{1}{6}\left[\frac{2}{b}+ c -\frac{3\xi^2}{2\lambda} \right].
\eeq
We showed how the effect of radiative corrections on the action 
could be analysed by considering the expansion  
\beq\label{eq:philoops}
\Gamma(\lambda_i,r,\rho/\mu) = S_{cl}(\lambda_i,r)
+ B(\lambda_i,r)\log(\rho/\mu) 
+ \frac{C(\lambda_i,r)}{2}\log^2(\rho/\mu) +\ldots,
\eeq
where $\rho=\sqrt{R},$  and the collection of dimensionless
coupling constants  $\{a,b,c, \xi,\lambda\}$ has been denoted
by $\lambda_i$.   The value of the effective
action for $\rho=\mu$ is simply the classical action.

In \reference{Einhorn:2014gfa}, we showed that the condition for an extremum 
corresponding to DT in this model (and others of this general form) is 
(to leading order) 
\beq\label{eq:b1os}
B_1^{(os)}=\sum_i\beta_{\lambda_i}\frac{\pa S_{os}}{\pa\lambda_i}=0,
\eeq
where $B_1^{(os)}$ is the ``on-shell'' one loop contribution to $B$, with 
``on-shell'' corresponding to $r=r_0$.
Such an extremum corresponds 
to a minimum if $\lambda > 0$ and
\beq
\varpi_2 =\half\left[C_2-{\left(B'_1\right)^2}/{S_{cl}^{''} }\right]\Big|_{r=r_0}>0,
\eeq
where $C_2$ is the on-shell leading (two-loop) contribution to $C$, 
and 
\beq
B'_1 = \frac{\partial}{\partial r}B_1(\lambda_i,r).
\eeq
Moreover, we used the RG to show that 
\beq\label{eq:varpi2}
\varpi_2=\half\left[\left(\beta_{\lambda_i}^{(1)}\frac{\partial}{\partial \lambda_i }
\right)^{\!2}\! S_{cl}
-\frac{1}{S_{cl}^{''}}\left(\!\beta_{\lambda_i}^{(1)}
\frac{\partial}{\partial \lambda_i } S'_{cl}(\lambda_i,r)
\!\right)^{\!2}  \right]\Big|_{r=r_0}.
\eeq
We find 
\begin{align}
\begin{split}
B_1^{(os)} &= \frac{1}{240x^2y^2}\big(1620x^4\xi^6+540x^4\xi^5+45x^4\xi^4-4320x^3\xi^4y\\
&-360x^3\xi^3y+60x^3\xi^2y 
+900x^2\xi^4+2880x^2\xi^2y^2+1800x^2\xi^2y\\
&-480x^2\xi y^2-826x^2y^2-2400x\xi^2y-2400xy^2+1600y^2\big).
\end{split}
\end{align}
For $\xi=0$, $B_1^{(os)}$ is independent of $y$:
\beq
B_1^{(os)} = -\frac{413x^2+1200x-800}{120x^2},
\eeq
and $B_1^{(os)}=0$ then has solutions $x= -3.465$, $x = 0.5591$.

\begin{figure}[tp]
\centering
\setlength\fboxsep{4pt}
\fbox{\includegraphics[width=4.5in]{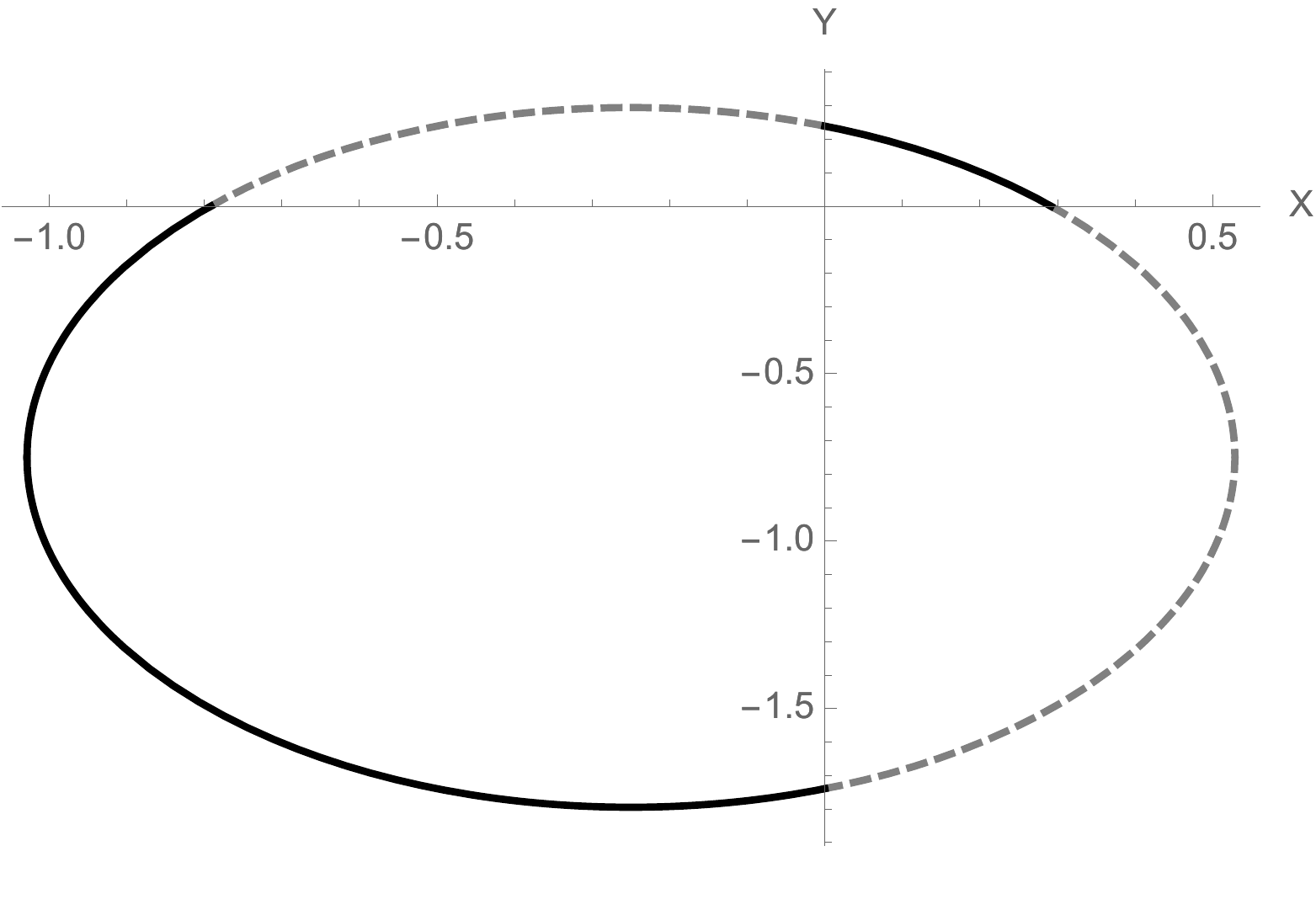}}
\caption{The $B_1 = 0$ ellipse.} \label{fig:1}
\end{figure}
We can write $B_1^{(os)}$ in terms of $z$, where $z \equiv 3x\xi^2/(4y)$:
\beq
B_1^{(os)} = (\frac{20}{3x^2} + 12\xi^2)(z-1)^2 + 2(\frac{5}{x}+\xi(2z+1))(z-1)
+\frac{1}{3}z(z+1)-\frac{413}{120},
\eeq
or in terms of $z'= z-1$, $\xi' = \xi + 1/6$:
\beq
B_1^{(os)} = z^{'2} (12\xi^{'2} +\frac{20}{3x^2})
+z' (6\xi' +\frac{10}{x}) -\frac{111}{40} 
\eeq
or
\beq
\label{eq:ellipse}
B_1^{(os)} = 12(X+\frac{1}{4})^2 +\frac{20}{3}(Y+\frac{3}{4})^2
-\frac{291}{40}, 
\eeq 
where $X = \xi^{'}z'$ and $Y = z'/x$. We thereby express 
$B_1^{(os)}$ in terms of two variables only, in terms of which the solutions 
to $B_1^{(os)} = 0$ lie on an ellipse, depicted in Fig.~1, 
enclosing the region
\beq
-1.0286 < X < 0.5286,\quad -1.7946 < Y < 0.2946.
\eeq
However, in order to obtain the correct sign for the 
Einstein term consequent to $\vev\phi\neq 0$ 
(and also by examination of the conformal scalar modes\cite{Einhorn:2014gfa}), 
we must require that  
$\xi > 0$ at the DT scale, corresponding to the constraint
\beq
\frac{X}{xY} > \frac{1}{6}.
\eeq
Since we have already concluded that $x > 0$ for 
all points in the UVFP catchment basin, 
$X$ and $Y$ must have the same signs. Thus, 
depending on the value of $x$, only portions of the
first and third  quadrants in Fig.~1 correspond to regions where
$\xi>0.$ The allowed range is depicted in Fig.~2a.

\begin{figure}[bp]
\centering
\begin{subfigure}[bp]{.49\textwidth}
\hskip-.25in
\centering
\setlength\fboxsep{2pt}
\fbox{\includegraphics[width=\textwidth]{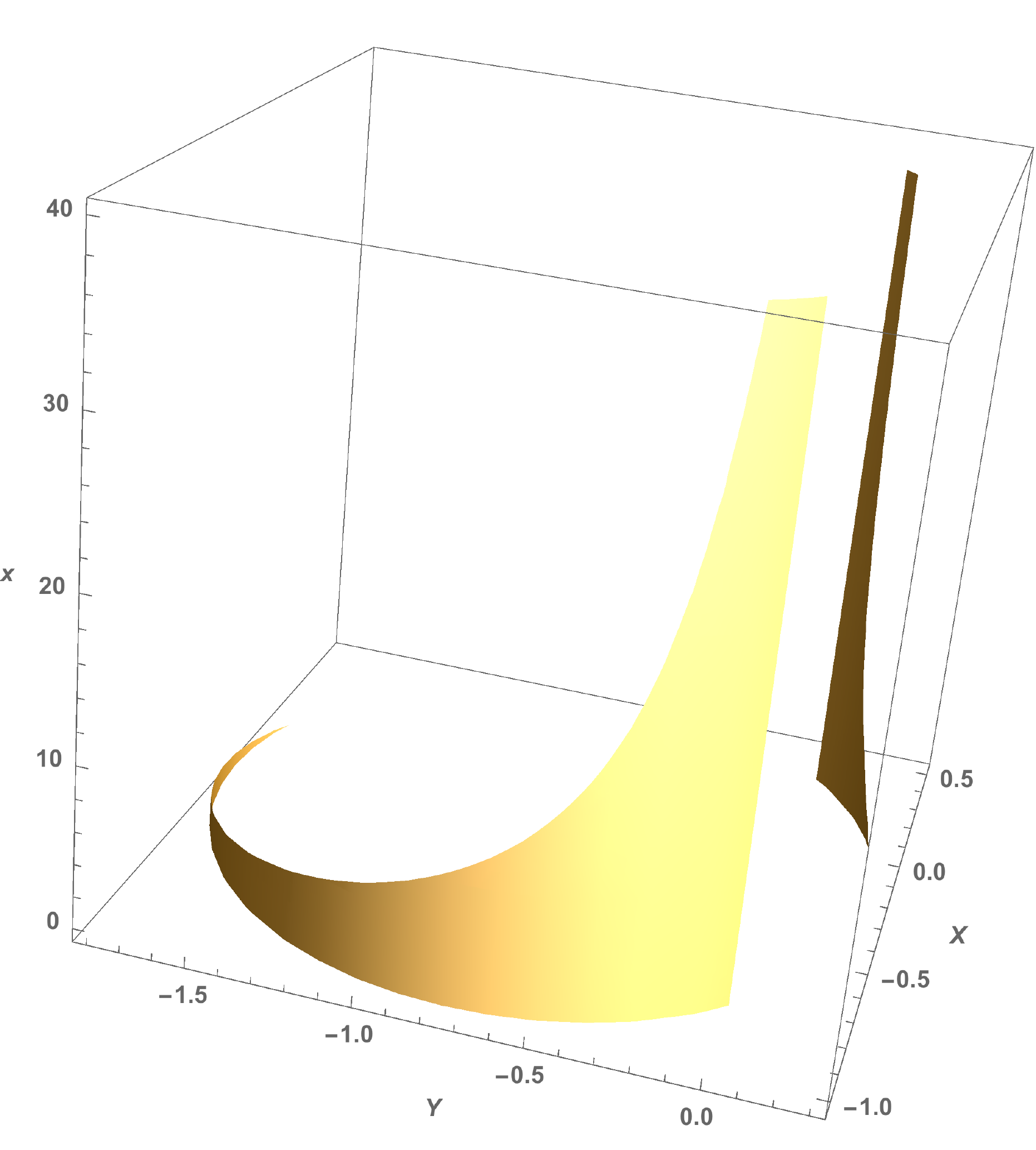}}
\caption{$\xi>0$.} \label{fig:2a}
\end{subfigure}
\begin{subfigure}[btp]{.48\textwidth}
\centering
\setlength\fboxsep{2pt}
\fbox{\includegraphics[width=\textwidth]{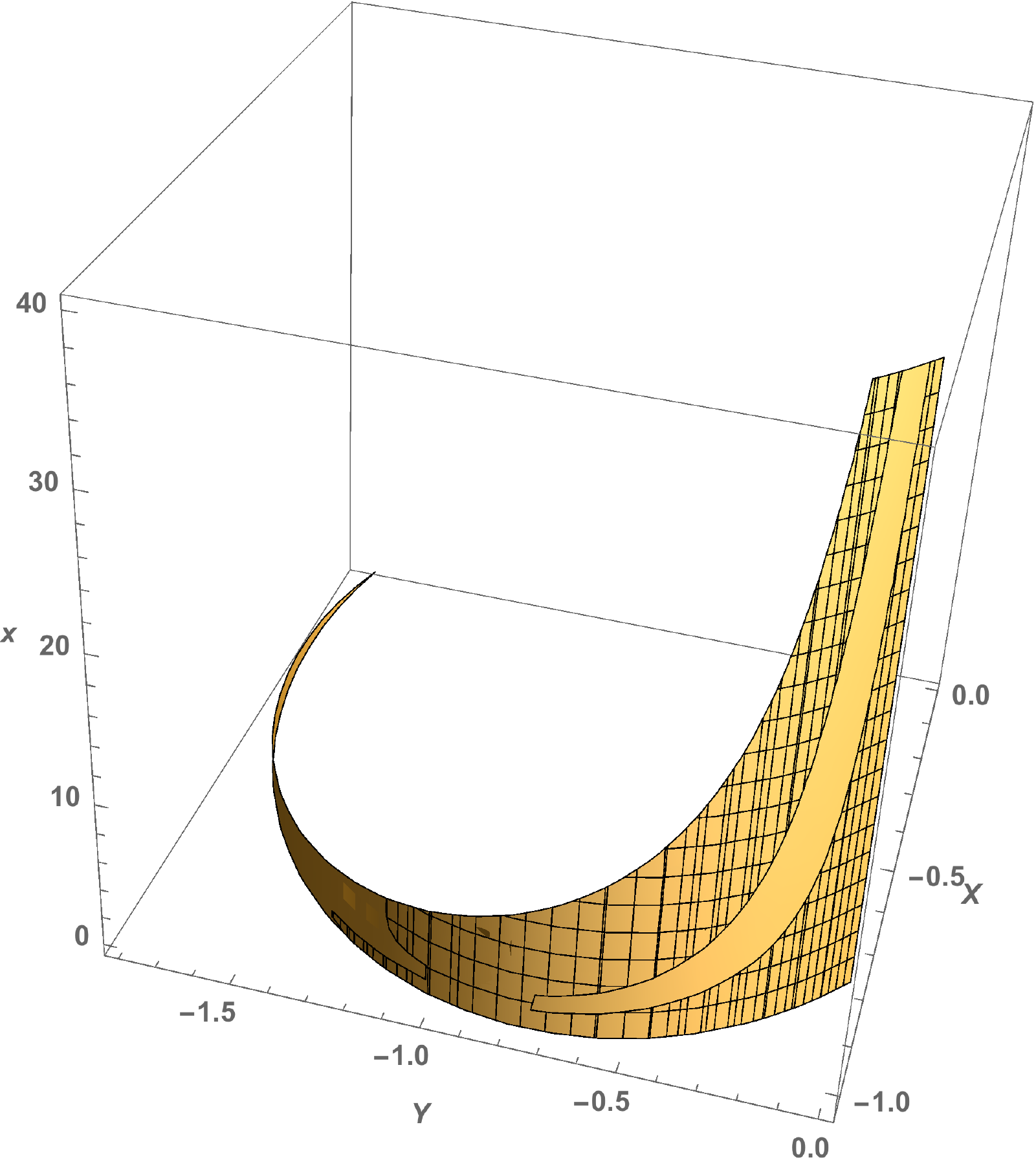}}
\caption{ $\xi, \varpi_2>0$.} \label{fig:2b}
\end{subfigure}
\caption{$B_1=0$ with constraints.}
\end{figure}

Similarly, $\varpi_2$ may be expressed in terms of $x,\xi$ and $z:$
\begin{align}
\label{eq:varpiz}
\begin{split}
\varpi_2 &= -a\big(10368x^4\xi^4z^4-23328x^4\xi^4z^3+6912x^4\xi^3z^4\\
&+7776x^4\xi^4z^2-6048x^4\xi^3z^3+1728x^4\xi^2z^4+12960x^4\xi^4z\\
&-9288x^4\xi^3z^2+1152x^4\xi^2z^3 +192x^4\xi z^4-7776x^4\xi^4+9072x^4\xi^3z\\
&-3636x^4\xi^2z^2+456x^4\xi z^3+8x^4z^4+4320x^3\xi^2z^3+11520x^2\xi^2z^4\\
&-648x^4\xi^3+648x^4\xi^2z-222x^4\xi z^2+34x^4z^3-8640x^3\xi^2z^2\\
&+1440x^3\xi z^3-34560x^2\xi^2z^3+3840x^2\xi z^4+108x^4\xi^2-84x^4\xi z\\
&+15x^4z^2+4320x^3\xi^2z-720x^3\xi z^2+120x^3z^3+34560x^2\xi^2z^2
-6240x^2\xi z^3\\
&+320x^2z^4 -720x^3\xi z+120x^3z^2-11520x^2\xi^2z
+960x^2\xi z^2 -80x^2z^3\\
&+330x^3z+1440x^2\xi z +13388x^2z^2+19984xz^3+3200z^4-13628x^2z \\
&-39968xz^2-9600z^3+19984xz+9600z^2-3200z\big)/(576x^3z),
\end{split}
\end{align}
or in terms of $(x,X,Y)$:
\begin{align}
\begin{split}
\label{eq:varpiXY}
\varpi_2 &= -a\Big(1296X^4\big(7 + 4xY\big) + 108X^3\big(65 + 44xY\big)\\ 
&+144X^2\big(13 + (15 + 11x)Y + (40 + 15x + x^2)Y^2 + 40xY^3\big)\\ 
&+ 3X\big(57 + 60(6 + x)Y + 4(220 + 90x + 3x^2)Y^2 + 880xY^3\big)\\
&+Y(1 + xY)\big(285 + 7094Y + 9992Y^2 + 1600Y^3\big) \Big)/\big(288Y(1 + xY)\big).    
\end{split}
\end{align}
Recall that, in order to have local stability (i.e., a positive dilaton
mass$^2$,) we must have $\varpi_2 > 0$.  Clearly, from \eqn{eq:varpiXY},
if $X$ and $Y$ are both positive, $\varpi_2<0,$ so  the first quadrant
in Fig.~1 is ruled out.  Therefore, since  $x>0,\xi>0,$ we must have
both $X$ and $Y$ negative.  
In Fig.~2b,  that portion of Fig.~2a having
$\varpi_2>0$ has been inscribed with a mesh. This is the region of
parameter space corresponding to DT that is locally stable with
attractive gravity.   Note that, unlike $B_1$, which has no explicit dependence on $a,$ $\varpi_2$ is explicitly proportional to $a$.  Further, if we were to restore the suppressed factors of $\kappa\equiv 1/(16\pi^2)$ in \eqns{eq:varpiz}{eq:varpiXY}, 
$\varpi_2$  would be preceded by a factor of $\kappa^2,$ as is to be expected for a two-loop correction, so that we can expect $\varpi_2\ll 1.$

For a consistent model, it must be that the values of the
coupling constants in this regime, when run from the DT scale up to
higher scales, approach the UVFP.
 This is a strong constraint and, in fact, fails for this model, as will be discussed in the next section.

\section{Basin of Attraction of the UVFP}

Although we have determined that the minimal model has a UVFP, we have
not delineated the basin of attraction of that point, i.e., the region
of all values of the renormalized coupling constants at finite scales
whose UV behavior approaches the UVFP.  In particular, in order to have
a complete theory, the values of the couplings where DT occurs $(B_1=0)$
must lie within this catchment basin.

In our earlier paper~\cite{Einhorn:2014gfa}, we showed rather easily
that DT occurs in a phase of the theory distinct from the UV catchment
basin.  With the SS beta-functions, we found a very different value for
the UVFP and a different equation for DT.  Nevertheless, by means of a
hopefully exhaustive exploration of numerical solutions of these
equations, we find that, even though there are regions of the  $B_1=0,$ 
$\xi>0$ surface that are locally stable $(\varpi_2>0)$  it apparently
remains true that these regions of DT  stability lie outside the UVFP
catchment basin. However, we should remark that the DT scale generally
lies well outside the neighborhood of the UVFP where a linear
approximation suffices, and we have not found a convincing {\it
analytical\/} argument in this nonlinear regime.

There is a further limitation to this conclusion associated with
the fact that we have neglected fermions and possible Yukawa couplings
with our scalar.   We shall discuss the inclusion of "sterile" fermions
in the next section, while remarking on the potential impact of Yukawa
interactions here. In the absence of gauge and gravitational couplings,
Yukawa couplings are never AF.  With the addition of renormalizable
gravity alone, Yukawa couplings $h_i$ are AF for small enough values at
the ``starting'' scale\footnote{For a review, see BOS, Sec.~9.5--9.6.
With the inclusion of gauge couplings, the situation becomes more
complicated; see \reference{EJ}.}.  In fact, they vanish even faster
than the gravitational coupling $a.$  However, as we proceed to  lower
scales, seeking a value where DT occurs, it is not clear that they
remain negligible.  Our discussion will continue to assume that they can
be ignored, but this ought to be explored further since fermions do
affect the RG flow of all the couplings and the additional equations
involving the Yukawa couplings make the determination of the RG-flows 
that much more challenging.  

Returning to the question of the RG flow, recall that the UVFP is at
$(x=x_0\approx39.8, \xi=0, y=0)$. In terms of the variables in Fig.~2,
this corresponds to $x\approx39.8,$ $X \approx -0.167,$ and $Y \approx
-0.025.$  The value of $x$ lies in the upper region shown in Fig.~2,
with $Y$ near $0$, and $X$ not far from its value at the center of the XY-cylinder ($X=-1/4$).
 This point lies far from the crosshatched regions shown in Fig.~2b, where locally
stable DT occurs.  The question is whether, starting near this UVFP and
running down to lower scales, the couplings intersect those regions.

First of all, one may not start just anywhere in a neighborhood of the
UVFP.  We argued in \reference{Einhorn:2014gfa} that the EPI converges
only if $a,x,\xi,y$ are all positive in the UV.  Moreover, we have shown 
above that  at the DT scale, we require $\xi > 0$ in order to generate 
Einstein-Hilbert gravity, and $a,x > 0$ in order to 
lie in the catchment basin of the UVFP.  The sign of $a$ cannot change in
perturbation theory, and it is positive and monotonically increasing as
one runs to lower scales. Since $x_0 \approx 39.8>0,$ any value of $x$
near there will do.  One can see from \eqn{eq:betabarmatC} that only the
term linear in $y$ is important near the UVFP, and its coefficient is
negative, as required for AF.  Thus, starting from an initial value
$y_0>0$, $y$ always increases as $u$ decreases (i.e., as $a$ increases.)
Stated otherwise, in flowing to lower scales, $y$ is always repelled from $0.$  
That is about all that can be said with certainty. 
The ratio $x$ can increase or decrease, depending on whether
it starts at a point greater or less than $x_0$.  To first order, $\xi$
may also increase or decrease depending on the sign of 
$(y-\xi(x_0-10/x_0)/3).$  Thus, even in linear approximation, the
behavior is complicated.  In the nonlinear regime relevant to DT, the interplay of the
different couplings is even harder to discern, and numerical studies
bear out that a variety of complicated trajectories can emerge.   We have also
explored various plots running toward larger $u$ (smaller $a$) starting
from points on the $B_1^{(os)}=0$ surface, where $\varpi_2>0.$ We have
found none that lead to the UVFP.

 We illustrate two varieties of behaviors of the running couplings in 
Fig.~3, both starting near the UVFP and running down toward the IR.
The starting values for each curve are given in their figure caption.
In Fig.~3a, $x,y,\xi$ all decrease toward the IRFP given in
the second row of Table~1. 
In Fig.~3b,  $x$  increases above $x_0,$ and,
if we continued following $y,$ we would see that it approaches a
singularity at negative $y,$ where perturbation theory breaks down. 
In both cases, after initially increasing, $y$ peaks and then decreases to
negative values of $y$.  

\begin{figure}[btp]
\centering
\begin{subfigure}[btp]{.45\textwidth}
\hskip-.4in
\centering
\setlength\fboxsep{2pt}
\fbox{\includegraphics[width=\textwidth]{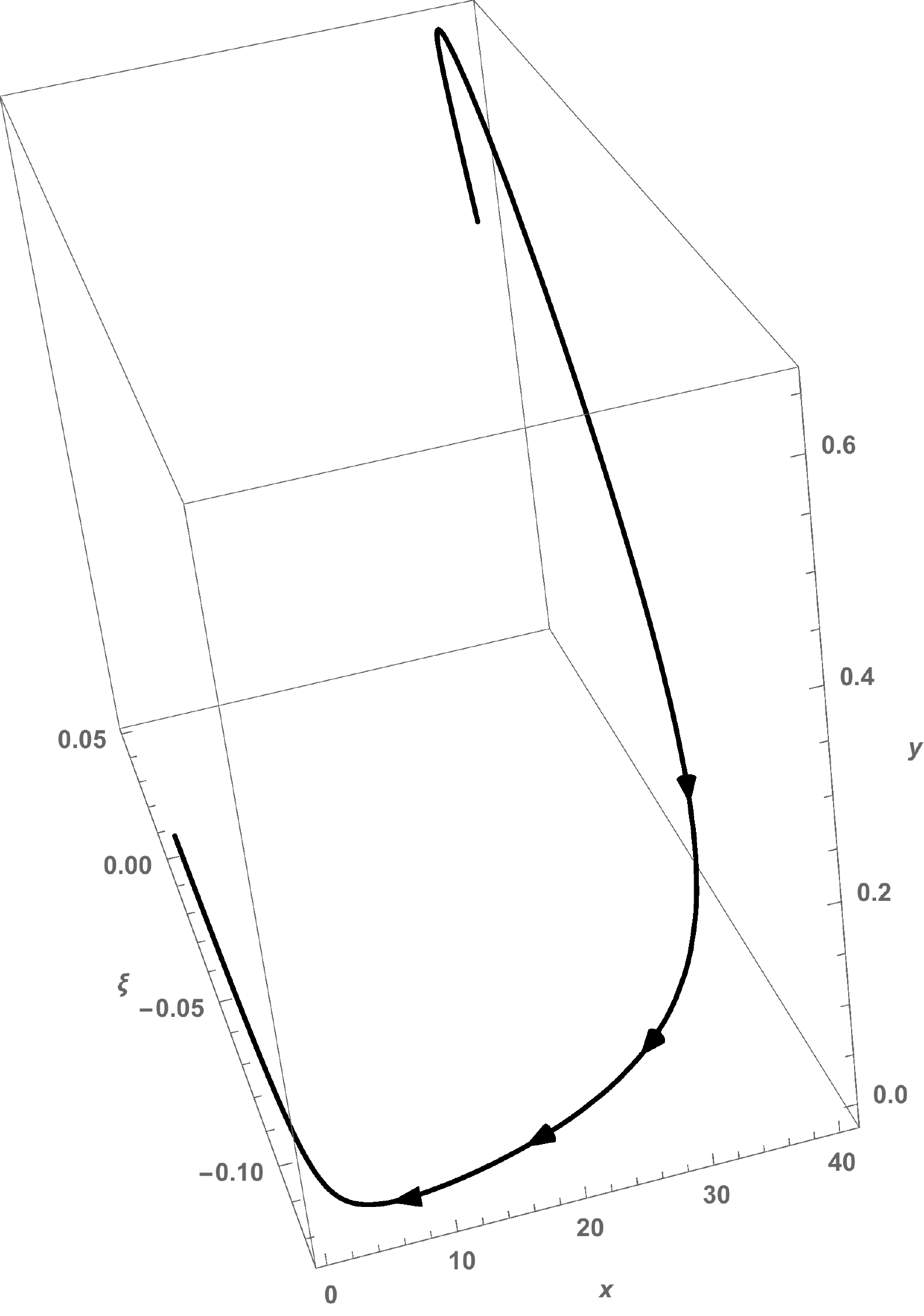}}
\caption{$\!x=38.03,\xi=0.0362,y=0.458$}  \label{fig:3a}
\end{subfigure}
\begin{subfigure}[btp]{.47\textwidth}
\centering
\fbox{\includegraphics[width=\textwidth]{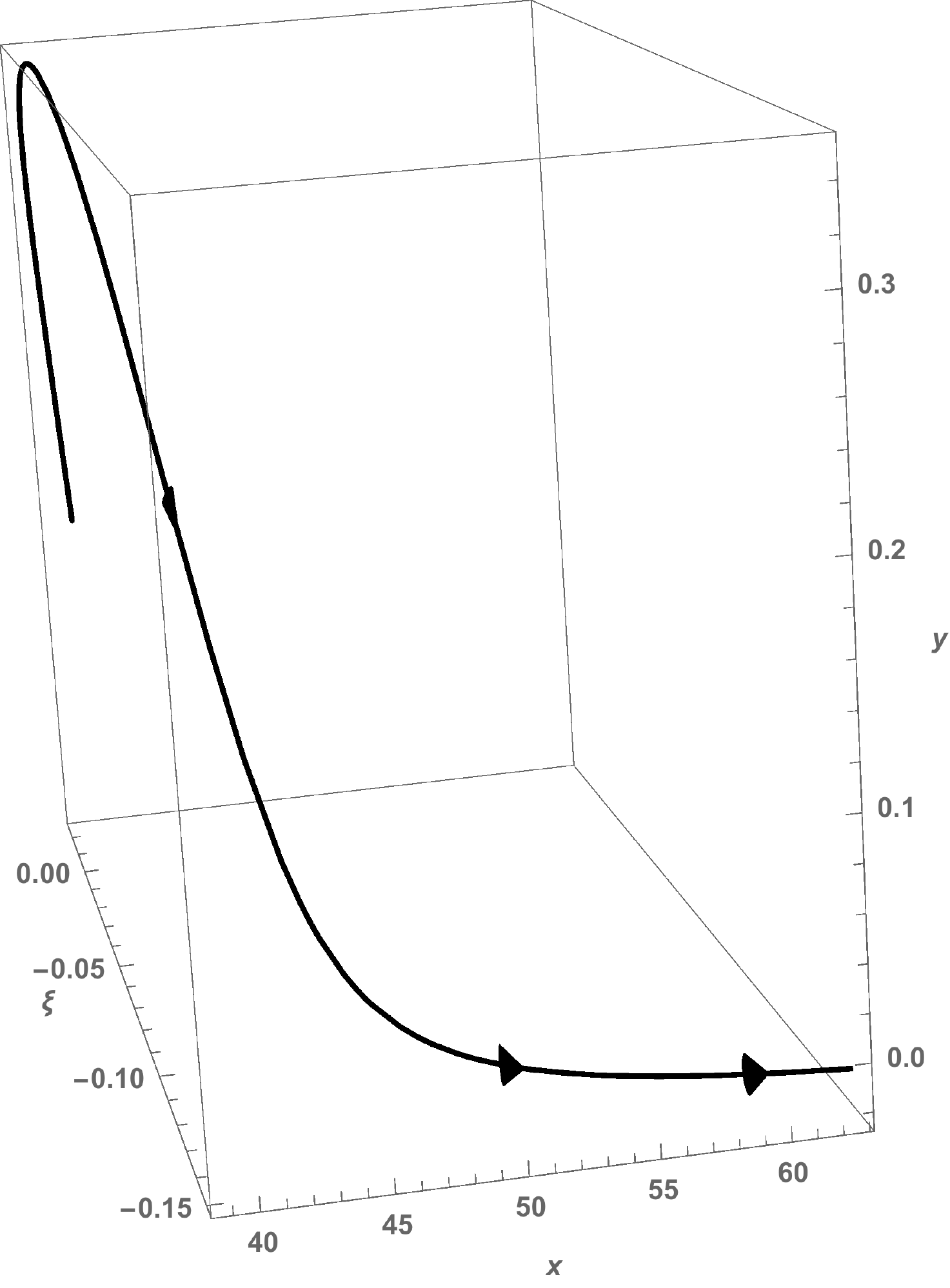}}
\caption{$\!x=39.27,\xi=0.0106,y=0.126$}  \label{fig:3b}
\end{subfigure}
\caption{Running couplings down from near the UVFP.}
\end{figure}

Our numerical explorations suggest the following conclusions.  In order
to have $B_1^{(os)}=0$ in a range where $\varpi_2>0,$ we must have  $y
\gtrsim 1.3.$  However, starting near the UVFP, it appears that $y$
increases initially but reaches a maximum value at some value of $y
\lesssim 0.8.$  That is not hard to believe, since the first and third
terms of \eqn{eq:betabarmatC} are positive for all values of the
couplings and beyond the linear regime, we tend to have
$\overline{\beta}_y>0.$  Even though $\overline{\beta}_y<0$ in the
linear regime, it can change sign rather quickly as $u$ decreases. 
Another way to see the challenge is to rewrite  $\overline{\beta}_y$ as
\beq
\overline{\beta}_y=18y^2 +\frac{1099}{60}y+\frac{5}{2}\xi^2+
x(6\xi+1)^2\left(x\frac{\xi^2}{8} -\half y\right).
\eeq
The only negative term is the last, so one can see the difficulties
sustaining  $\overline{\beta}_y<0$ beyond linear approximation, but
exactly how large it can get depends on the starting values of $x$ and
$\xi$ and their running. In Fig. 3, we chose examples where the 
 increase of $y$ is relative large, but it turns around long before it approaches 
 $y\approx 1.3.$
 
We conclude that the catchment basin of the UVFP describes a phase of the
theory demarcated from regions where locally stable DT occurs. However, 
there are regions of parameter space with $a,b,\xi,y$ all positive, where DT 
occurs at a scale which we may associate with the Planck mass via 
$M_P^2 \sim \xi\vev\phi^2$. The effective field theory below this scale is 
Einstein Gravity with a massive dilaton.

\section{Including Fermions}

In view of the negative conclusion of the previous section it is
worthwhile considering  modifying  the minimal model by including 
additional matter fields. The simplest possible such generalisation
would involve the inclusion of such fields without  additional
dimensional couplings. This could clearly be done in a natural way  by
invoking a global symmetry with respect to which the scalar $\phi$
transformed as a singlet,  and adding a fermion multiplet without a
quadratic invariant with respect to this symmetry. Under these
conditions, there can be no Yukawa couplings, so that the only changes to
our calculations would be to alter $b_1$ and $b_2$ (see \eqn{eq:bgb2b1}).  

For general $b_1$, $b_2$ , the reduced beta-function 
\eqn{eq:betabarmatA}-\eqn{eq:betabarmatC} become
\begin{subequations}
\begin{align}
\frac{dx}{du}\equiv\overline{\beta}_x&=
-\frac{10}{3} + (5+ b_2)x 
-\frac{1}{24} x^2 \left(10+(1+6 \xi)^2\right));
\label{eq:betabarmatFA}\\
\frac{d\xi}{du}\equiv\overline{\beta}_{\xi}&=\left(6\xi 
+ 1\right)y +\frac{\xi}{6}\left(\frac{20}{x}-x(6\xi+1)(3\xi+2)\right);
\label{eq:betabarmatFB}\\
\label{eq:betabarmatFC}
\frac{dy}{du}\equiv\overline{\beta}_y&=18y^2 
+ y\left(5+b_2-\half x (1+6\xi)^2\right)
+\frac{\xi^2}{8}(20+(6\xi+1)^2x^2).
\end{align}
\end{subequations}
where with the addition of a fermion multiplet we now have 
\beq
N_a = \frac{1}{60}\left(1+3N_F\right)\quad \hbox{and}\quad
N_c = \frac{1}{360}\left(1+\frac{11}{2}N_F\right).
\eeq
Note that \eqn{eq:betabarmatFB} is unchanged. 

It is possible to find the resulting FPs for general $N_F$, but the resulting expressions 
are unwieldy. However, the FP corresponding to the UVFP in Table~\ref{table:FP}
becomes (for general $b_{1,2}$):
\beq
x_{FP} = \frac{1}{11}\left(60 + 12b_2 + 4\sqrt{9b_2^2  + 90 b_2 + 170}\right),
\quad \xi = y = 0 
\eeq
or for case of the fermion multiplet: 
\beq
x_{FP} = \frac{1}{55}\left(1099 + 3N_F + \sqrt{1185801+6594 N_F + 9N_F^2}\right), 
\quad \xi = y = 0.
\eeq
It is straightforward to show that this FP is UV attractive for arbitrary $N_F \geq 0$, 
or indeed arbitrary $b_2 > 133/10$. 

The result for $B_1^{(os)}$ (\eqn{eq:ellipse}) becomes 
\bea
\label{eq:ellipseN}
B_1^{(os)} &=& 12(X+\frac{1}{4})^2 +\frac{20}{3}(Y+\frac{3}{4})^2
-\frac{35}{12}-b_1\nn 
 &=& 12(X+\frac{1}{4})^2 +\frac{20}{3}(Y+\frac{3}{4})^2
-\frac{291}{40}- \frac{11}{720}N_F.
\eea 
Note that $B_1^{(os)}$ depends on  $b_1$, that is on the beta-function
for the coefficient of the Gauss-Bonnet term; 
so as we remarked in  section~\ref{sec:FP}, 
ignoring this term 
is not correct, even for $N_F = 0$. \eqn{eq:ellipse}\ is 
replaced by  \eqn{eq:ellipseN}\  
and we see that ignoring $b_1$ would make a significant numerical difference.

However the formulae for $\varpi_2$, \eqns{eq:varpiz}{eq:varpiXY}, do not change if we 
keep $b_1$ general, but they do depend on $b_2$:
\bea
\varpi_2 &=& -\frac{a}{288Y(1 + xY)}
\Big(1296X^4\big(7 + 4xY\big) + 108X^3\big(65 + 44xY\big)\nn 
&+& 144X^2\big(13 + (15 + 11x)Y + (40 + 15x + x^2)Y^2 + 40xY^3\big)\nn 
&+& 3X\big(57 + 60(6 + x)Y + 4(220 + 90x + 3x^2)Y^2 + 880xY^3\big)\nn
&+& 5Y(1 + xY)\big(57+4 (18 b_2+ 115) Y+48 (2 b_2 + 15\,) Y^2+320 Y^3\big) \Big).
\eea
We see that the property that local stability requires both $X$ and $Y$ negative 
is sustained by this generalisation.

\section{Conclusions}

We have shown that the theory consisting of  renormalisable $R^2$
quantum gravity coupled to a single scalar field in  a scale-invariant
way undergoes dimensional transmutation in a  manner which can be
credibly described by perturbation theory. Below the DT scale, the 
theory describes Einstein gravity coupled to a scalar dilaton which
obtains a mass  through spontaneous breaking of scale invariance. 
We also found that  the theory possesses an Ultra-Violet Fixed Point for
coupling ratios, such that  all the couplings tend to zero as this FP is
approached, with the ratio  $x = b/a \to 39.8$. Since $a > 0$ is
required for Asymptotic Freedom, it follows that $b > 0$  in the
neighbourhood of the FP. In fact in \reference{Einhorn:2014gfa}\ we
argued that  both $a,b > 0$  (and $\lambda > 0$) are  
required for convergence of the EPI, so
the theory  is well behaved in the UV for couplings in the FP catchment
basin. It should be noted that here we appear to differ from  SS, who in our
notation seem to require $a > 0$ but $b < 0$.

However, although the region of parameter space for the dimensionless
couplings where DT occurs includes a region with $x,y > 0$ and also $\xi > 0$, 
which we require to generate Einstein gravity, 
it turns out that the theory becomes strongly coupled at  higher
scales, with couplings approaching Landau poles.  Thus this particular
theory is not an ultraviolet  (UV) complete theory of Einstein gravity.
This is a disappointing  outcome since there {\it are\/} regions of
parameter space where all the couplings are asymptotically free, with
coupling constant ratios approaching the  UV Fixed Point.

Although we have adopted the beta-functions of SS, we wish to emphasize
that our results differ from theirs in significant ways. Their
determination of the analog of our function $B_1$ would omit any
contribution from the Gauss-Bonnet term.  Moreover, their
criteria for  determination of the DT scale involves an approximation
that differs significantly from ours.

Further, we determined the  two-loop value of the dilaton
mass$^2$ in order to ascertain whether the DT extrema are locally
stable.  Finally, we explored the basin of attraction of the UVFP,
showing via the renormalization group that there are apparently no paths
in this catchment that, at lower scales, undergo DT in a manner that
satisfies the physical constraints.  We therefore regard their
applications of these sorts of classically scale-invariant models 
somewhat skeptically. 

In a subsequent paper~\cite{EJ}\ we will extend our formalism to Grand
Unified Theories, where we show that once again it is possible to
construct completely Asymptotically  Free models, with coupling constant 
ratios approaching fixed points. It transpires that to achieve this it
is necessary  to add enough matter fields to make the one loop gauge
beta-function coefficient  as numerically small as possible (while,
obviously, remaining negative). This was demonstrated long ago in flat
space~\cite{Cheng:1973nv} and remains true in the presence  of 
gravitational corrections~\cite{Buchbinder:1989ma, Buchbinder:1989jd}. 

It is also possible to exhibit GUT models which undergo Dimensional
Transmutation  in the same manner as we have described here.  Moreover,
by appropriate choice of scalar representation it is possible to arrange
that the same scalar vacuum expectation value generated by DT both
produces Einstein gravity {\it and\/} breaks the Grand Unified symmetry.
The crucial question (which had a disappointing answer in the model of
this paper) is whether  there are DT regions of parameter space in the
catchment basin of a UVFP.  We will answer this question in
\reference{EJ}, where we construct a model based on the gauge group SO(10) 
with an adjoint scalar representation.  This scalar acquires a vev via DT, 
breaking  the SO(10) symmetry so as to leave unbroken
the maximal subgroup SU(5)$\otimes$U(1). Moreover, we have shown that 
there is a region of parameter space where DT occurs that satisfies 
all our requirements (such as generation of a ``right-sign'' Einstein term) 
{\it and\/} is in the catchment of a UV fixed point such that all the 
couplings are asymptotically free.  We thus have the basis for a UV complete extension of the Standard Model. 

Of course problems remain to be solved, not least of which being
the origin of the electroweak scale.  There is also the issue of the
(doubtful) unitarity of $R^2$ gravity, in both the minimal model
considered here and in the gauge theory extensions.  We discussed this
problem briefly in \reference{Einhorn:2014gfa} and will do so again
\reference{EJ}; suffice to say for now that we believe it is possible
that the combination of AF (at high energies) with DT as we run towards
the IR (so that if DT did not occur the theory would become strongly
coupled) leads to its solution.

\acknowledgments
We would like to thank A.~Salvio and A.~Strumia for helpful correspondence.  
DRTJ thanks  KITP (Santa Barbara), the Aspen
Center for Physics and CERN for hospitality and financial support. This
research was supported in part by the National Science Foundation under
Grant No. PHY11-25915 (KITP) and Grant No. PHY-1066293 (Aspen), and by the Baggs bequest (Liverpool).

\end{document}